\documentclass[aps,prx,reprint,superscriptaddress,longbibliography]{revtex4-1}
\usepackage[english]{babel}
\usepackage{amsmath}
\usepackage{amsfonts}
\usepackage{amssymb}
\usepackage{graphicx}
\usepackage{bm}
\usepackage{bbm}
\usepackage{braket}

\begin{document}

\title{Double Quantum Dot Floquet Gain Medium}
\author{J. Stehlik}
\affiliation{Department of Physics, Princeton University, Princeton, New Jersey 08544, USA}
\author{Y.-Y. Liu}
\affiliation{Department of Physics, Princeton University, Princeton, New Jersey 08544, USA}
\author{C. Eichler}
\affiliation{Department of Physics, Princeton University, Princeton, New Jersey 08544, USA}
\author{T. R. Hartke}
\affiliation{Department of Physics, Princeton University, Princeton, New Jersey 08544, USA}
\author{X. Mi}
\affiliation{Department of Physics, Princeton University, Princeton, New Jersey 08544, USA}
\author{M. J. Gullans}
\affiliation{Joint Quantum Institute, National Institute of Standards and Technology, Gaithersburg, Maryland 20899, USA}
\affiliation{Joint Center for Quantum Information and Computer Science,
	University of Maryland, College Park, Maryland 20742, USA}
\author{J. M. Taylor}
\affiliation{Joint Quantum Institute, National Institute of Standards and Technology, Gaithersburg, Maryland 20899, USA}
\affiliation{Joint Center for Quantum Information and Computer Science,
	University of Maryland, College Park, Maryland 20742, USA}
\author{J. R. Petta}
\affiliation{Department of Physics, Princeton University, Princeton, New Jersey 08544, USA}
%\pacs{03.67.Lx, 73.63.Kv, 85.35.Gv}
% 03.67.Lx - Quantum computation architectures and implementations
% 73.63.Kv - electronic transport in QD
% 85.35.Gv - single electron devices

\begin{abstract}
Strongly driving a two-level quantum system with light leads to a ladder of Floquet states separated by the photon energy. Nanoscale quantum devices allow the interplay of confined electrons, phonons, and photons to be studied under strong driving conditions. Here we show that a single electron in a periodically driven DQD functions as a ``Floquet gain medium," where population imbalances in the DQD Floquet quasi-energy levels lead to an intricate pattern of gain and loss features in the cavity response. We further measure a large intra-cavity photon number $n_{\rm c}$ in the absence of a cavity drive field, due to equilibration in the Floquet picture. Our device operates in the absence of a dc current -- one and the same electron is repeatedly driven to the excited state to generate population inversion.  These results pave the way to future studies of non-classical light and thermalization of driven quantum systems.
\end{abstract}

\maketitle

\section{Introduction}

A qubit coupled to a microwave resonator allows the study of fundamental light-matter interactions at the level of single photons \cite{cQED}.  The circuit quantum electrodynamics (cQED) architecture enables the exploration of a variety of phenomena. Measurements of the transmission through a superconducting cavity have been used to read out both superconducting \cite{strongCouplingWallraff} and spin qubits \cite{KarlFancy}. Qubits separated by relatively large distances have been coherently coupled using a cavity bus \cite{CoherentStorage,Majer2007}.  Finally, cQED enables the generation of classical and non-classical light \cite{Houck,astafiev,YinyuScience,HofheinzNature}. 

Light sources based on the cQED architecture generally use two different approaches. The first approach relies on coherently transfering an excitation from a qubit to a cavity. Early work demonstrated single photon generation by applying a $\pi$-pulse to a qubit to drive a transition from the ground state to the excited state. The qubit was then brought into resonance with the cavity for a short period of time, thereby transfering the excitation from the qubit to the cavity \cite{Houck}. $n$-photon Fock states were generated by repeating this process many times  \cite{HofheinzNature}. A second approach utilizes a source-drain bias to drive a current through a Cooper pair box or a double quantum dot (DQD) \cite{Hofheinz2011,YinyuPRL}. Here the source-drain bias continually repumps the excited state of the artificial atom, leading to population inversion and microwave frequency photoemission. For sufficiently high pumping rates a transition to above-threshold masing has been achieved \cite{astafiev,YinyuScience}.   

When qubits are strongly driven, intriguing quantum effects emerge, including Landau-Zener-St\"uckelberg interference \cite{ShevchenkoReview,Grifoni}, lasing without inversion \cite{AWI} and vacuum squeezing \cite{Kimble77,Schulte15}. The combination of superconducting resonators and semiconductors -- in the form of DQDs -- provides a versatile, electrically tunable quantum system that interacts strongly with light \cite{KontosCoupling,cavityDQDWallraff1,YinyuScience}.   

\begin{figure*}
	\begin{center}
		\includegraphics[width=1.75\columnwidth]{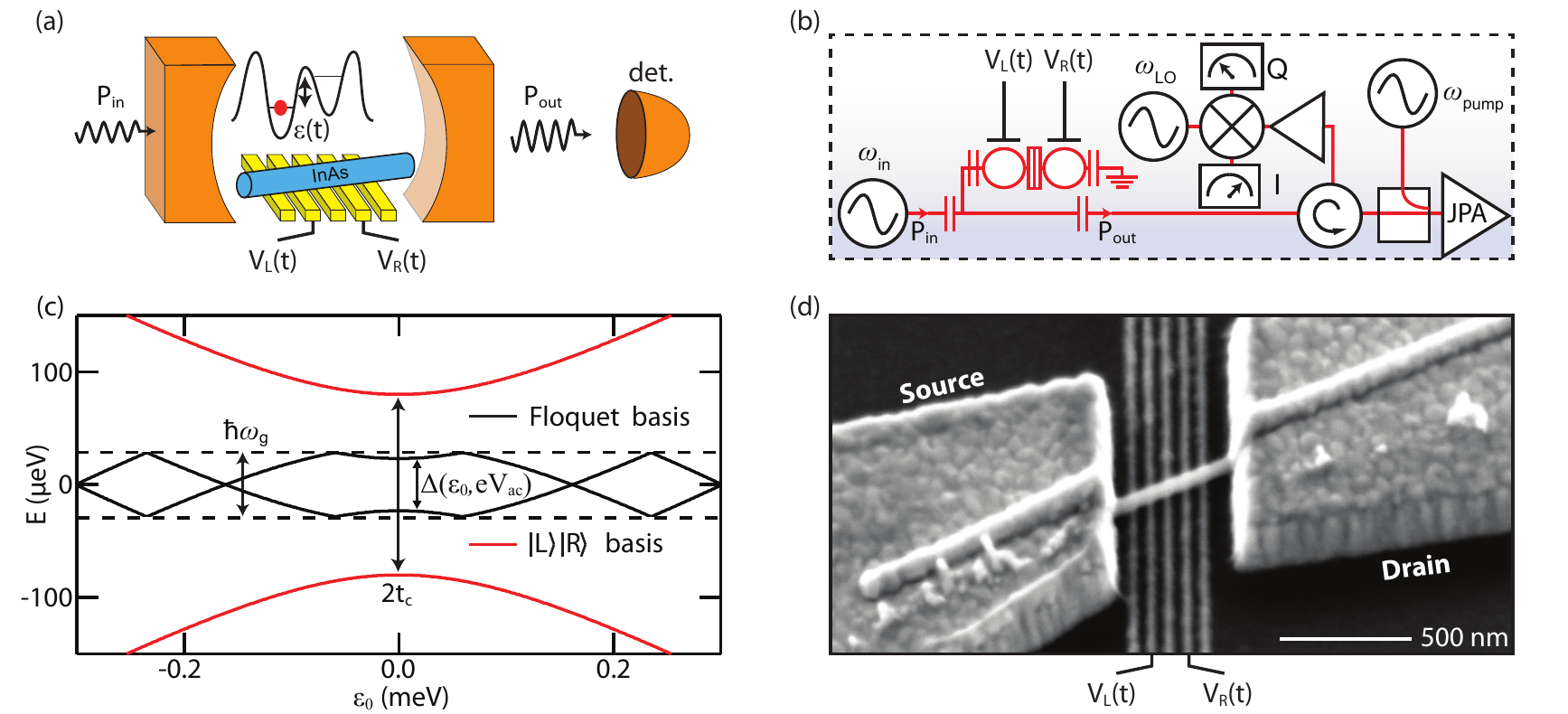}
		\caption{ (a) Driven double quantum dot photon source. The gain medium consists of a DQD whose energy level detuning $\epsilon(t)$ is driven periodically. A cavity containing the DQD is probed using a weak input signal with power $P_{\rm in}$. The signal exiting the cavity with power $P_{\rm out}$ is incident on a detector. (b) Circuit diagram illustrating the DQD, cavity, and linear amplification chain. (c) DQD energy level diagram, with the bonding and antibonding charge states shown in red. Interdot tunnel coupling results in a level splitting of $2t_c$ at $\epsilon_0$ = 0. In the Floquet basis the energy level spectrum is only defined modulo $\hbar \omega_g$, resulting in folding of the DQD levels. The splitting of the folded levels is $\Delta(\epsilon_0, eV_{\rm ac})$. (d) Scanning electron microscope image of the sample.  A single InAs nanowire is placed across five bottom gates and electrically contacted by source and drain electrodes. The detuning parameter is driven by applying oscillating voltages $V_{\rm L}(t)$ and $V_{\rm R}(t)$ to two of the gates.}
		\label{fig:sisyphus:setup}
	\end{center}	
	\vspace{-0.6cm}
\end{figure*}

In this paper we show that a periodically driven single electron confined to a cavity-coupled DQD functions as a ``Floquet gain medium," where population imbalances in the DQD Floquet quasi-energy levels lead to an intricate pattern of gain and loss features in the cavity response \cite{ShevchenkoReview,Grifoni,gullansSisyphus}. The operating regime of the device is distinct from previous work on voltage-biased Josephson junctions and semiconductor DQDs, where a relatively large current flow was required to achieve photoemission \cite{YinyuPRL}. Instead, our device is operated in Coulomb blockade, such that the net current flow through the device is negligible. One and the same electron is repeatedly driven to the excited state to generate population inversion. The drive field modifies the effective qubit level-diagram and leads to the formation of a ladder of Floquet bands, as shown in Fig.\ \ref{fig:sisyphus:setup}(c) \cite{PhysRevA.7.2203}. Periodic driving of the level detuning results in an effective population inversion that depends sensitively on the drive parameters and band alignment. An intricate photon emission pattern emerges as a result of cavity coupling, periodic driving, and strong electron-phonon coupling  \cite{Neilinger}.

\section{Driven DQD Gain Medium}

Our experimental setup is illustrated in Figs.\ \ref{fig:sisyphus:setup}(a--b) and consists of a half-wavelength ($\lambda/2$) microwave cavity that contains a DQD. The cavity has a center frequency $\omega_c / 2\pi$  = 7.553 GHz and linewidth $\kappa / 2 \pi$ = 1.5 MHz, resulting in a quality factor $Q$ $\approx 5\,000$. A single InAs nanowire DQD is placed at an anti-node of the cavity electric field  [see SEM image in Fig.\ \ref{fig:sisyphus:setup}(d)] \cite{KarlFancy}. The DQD confinement potential is generated by five gate electrodes that are located beneath the nanowire \cite{RevModPhys.75.1}.  The left $V_{\rm L}$ and right $V_{\rm R}$ gate voltages are used to tune the chemical potential of the left dot $\mu_{\rm L}$ and right dot $\mu_{\rm R}$, thereby setting the interdot detuning $\epsilon(t) = \mu_{\rm R} - \mu_{\rm L}$.  Measurements are taken near an interdot charge transition, where a single excess electron occupies either the left or right quantum dot.

To enhance the electric dipole coupling of this electron to the cavity electric field, the nanowire source contact is connected to the cavity center pin, resulting in a charge-cavity coupling rate $g_0 / 2 \pi \approx 80$ MHz. We probe the cavity by weakly driving the input port of the cavity at frequency $\omega_{\rm in} / 2 \pi$ and power $P_{\rm in} \approx -120$~dBm, see Fig.\ \ref{fig:sisyphus:setup}(a).  %We detect the radiation emitted from the cavity by first amplifying it with a Josephson parametric amplifier  \cite{CavesPRD1982,CavesPRA2012,LehnertJPA}, which is followed by a 4 Kelvin noise temperature high-electron mobility transistor amplifier and room temperature amplifiers. The amplified signal is demodulated using a field programmable gate array (FPGA) based heterodyne detection stage (see Appendix \ref{appendix:methods}).
We detect the radiation emitted from the cavity using a multistage amplification process starting with a nearly quantum limited Josephson parametric amplifier (see Appendix \ref{appendix:methods} for a detailed description of the experimental setup) \cite{CavesPRD1982,PhysRevA.39.2519,LehnertJPA}.

The InAs nanowire DQD is described by a charge qubit Hamiltonian \cite{PettaSeminal,StehlikPRB}:
\begin{equation}
\label{eqn:hdqd}
H_{\rm DQD} = \frac{ \epsilon(t)}{2} \sigma_{\rm z} + t_{\rm c}\sigma_{\rm x}. 
\end{equation}
Here $\sigma_{\rm x}$ and $\sigma_{\rm z}$ are the Pauli operators in the left--right charge basis and $t_{\rm c}$ = $80\pm10$~$\mu$eV is the interdot tunnel coupling. For $\epsilon_0 < 0$ the excess electron resides in the right dot (denoted $\ket{R}$) in the absence of any microwave excitation. The DQD is periodically driven at frequency $\omega_{\rm g} / 2 \pi$ by applying oscillating voltages $V_{\rm L}$ and $V_{\rm R}$ to the gates. The relative amplitude and phase of these voltages are set such that \cite{RocheCP}:
\begin{equation}
\epsilon(t) = \epsilon_0 + e V_{\rm ac} \sin(\omega_{\rm g} t).
\end{equation}
Here $e$ is the electron charge, $V_{\rm ac}$ is the amplitude of the microwave drive, and $\epsilon_0$ is the offset detuning (see Appendices \ref{appendix:methods:DQD} and \ref{appendix:methods:detuning} for details). 

\begin{figure*}
	\begin{center}
		\includegraphics[width=1.75\columnwidth]{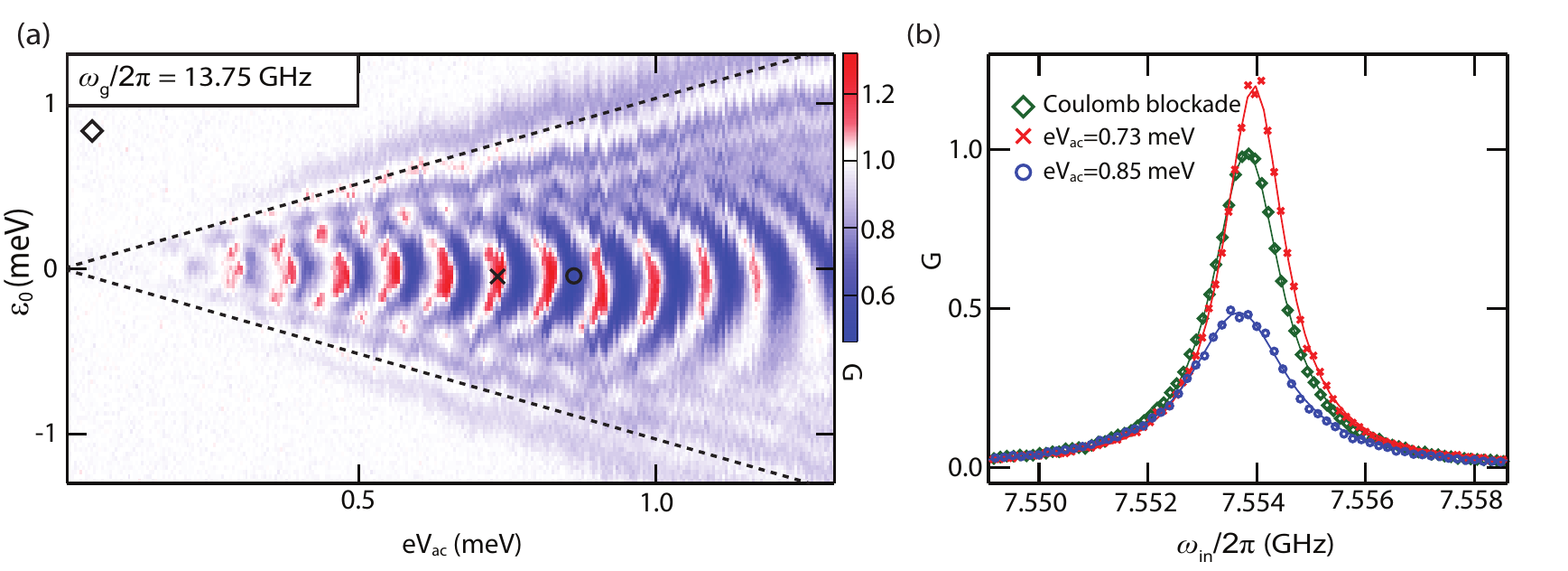}
		\caption{(a) Floquet gain medium interference pattern. The cavity power gain $G$ plotted as a function of offset detuning $\epsilon_0$ and ac drive amplitude $e V_{\rm ac}$. A pattern of alternating gain ($G$ $>$ 1, red regions) and loss ($G$ $<$ 1, blue regions) is observed. (b) $G$ plotted as a function of $\omega_{\rm in}/2\pi$ with: the DQD in Coulomb blockade (green diamond), with the DQD drive configured for loss ($e V_{\rm ac} = 0.85$ meV, blue circle), and with the DQD drive configured for gain ($e V_{\rm ac} = 0.72$ meV, red cross).  Solid lines are Lorentzian fits to the data.}
		\label{fig:sisyphus:transmission}
	\end{center}
	\vspace{-0.6cm}
\end{figure*}

We measure the cavity power gain $G$ in the presence of the periodic drive. The power gain is defined as $G = C P_{\rm out} / P_{\rm in}$, where $P_{\rm out}$ is the cavity output power and $C$ is a normalization constant set such that $G=1$ with the device configured in Coulomb blockade \cite{YinyuScience,YinyuPRL}. In Fig.\ \ref{fig:sisyphus:transmission}(a) we plot $G$ as a function of $e V_{\rm ac}$ and $\epsilon_0$ for $\omega_g / 2 \pi  = 13.75$ GHz.  The cavity gain data exhibit a distinctive interference pattern, with oscillations between regions of cavity gain ($G$ $>$ 1) and cavity loss ($G$ $<$ 1), as $eV_{\rm ac}$ and $\epsilon_0$ are varied. The boundary of this interference pattern has a ``V" shape that traces out the region where $|\epsilon_0| \sim e V_{\rm ac}$. Inside this region the ac drive sweeps the level detuning through $\epsilon$ = 0 and the gain displays an interference pattern \cite{astafiev}. Outside this region the transmission through the cavity is decoupled from the DQD dynamics and $G$ = 1.  
%Inside this region $G$ features strong interference pattern.

With $\epsilon_0$ = 0, the interference pattern shown in Fig.\ \ref{fig:sisyphus:transmission}(a) exhibits striking oscillations in $G$ as a function of $e V_{\rm ac}$. To further investigate these oscillations we measure the cavity gain as a function of input frequency $\omega_{\rm in} / 2 \pi$ [Fig.\ \ref{fig:sisyphus:transmission}(b)] at locations in the interference pattern where there is gain and loss. The conditions $\epsilon_0$ = 0 and $e V_{\rm ac} = 0.73$ meV correspond to a location where $G$ = 1.2, indicating that the Floquet gain medium is effectively inverted. Here the cavity gain is well-fit by a Lorentzian with a full-width at half-maximum (FWHM) of 1.2 MHz, significantly narrower than the bare cavity linewidth $\kappa/2\pi$ = 1.5 MHz. Alternatively, when we configure the driving parameters to a region of microwave loss ($\epsilon_0$ = 0, $e V_{\rm ac} = 0.85$ meV), the cavity gain is noticeably broader with a FWHM = 2.1 MHz and peak gain of 0.49, indicating additional loading of the cavity. Here the Floquet gain medium is not inverted. In comparison, the green curve in Fig.\ \ref{fig:sisyphus:transmission}(b) shows the cavity gain in the region outside of the interference pattern (where $|\epsilon_0| \gg e V_{\rm ac}$). In this region the FWHM = 1.5 MHz is equal to the bare cavity linewidth, indicating that the cavity is not being affected by the charge dynamics of the DQD.

\section{Photon emission in the absence of a current}

\begin{figure*}
\begin{center}
\includegraphics[width=1.75\columnwidth]{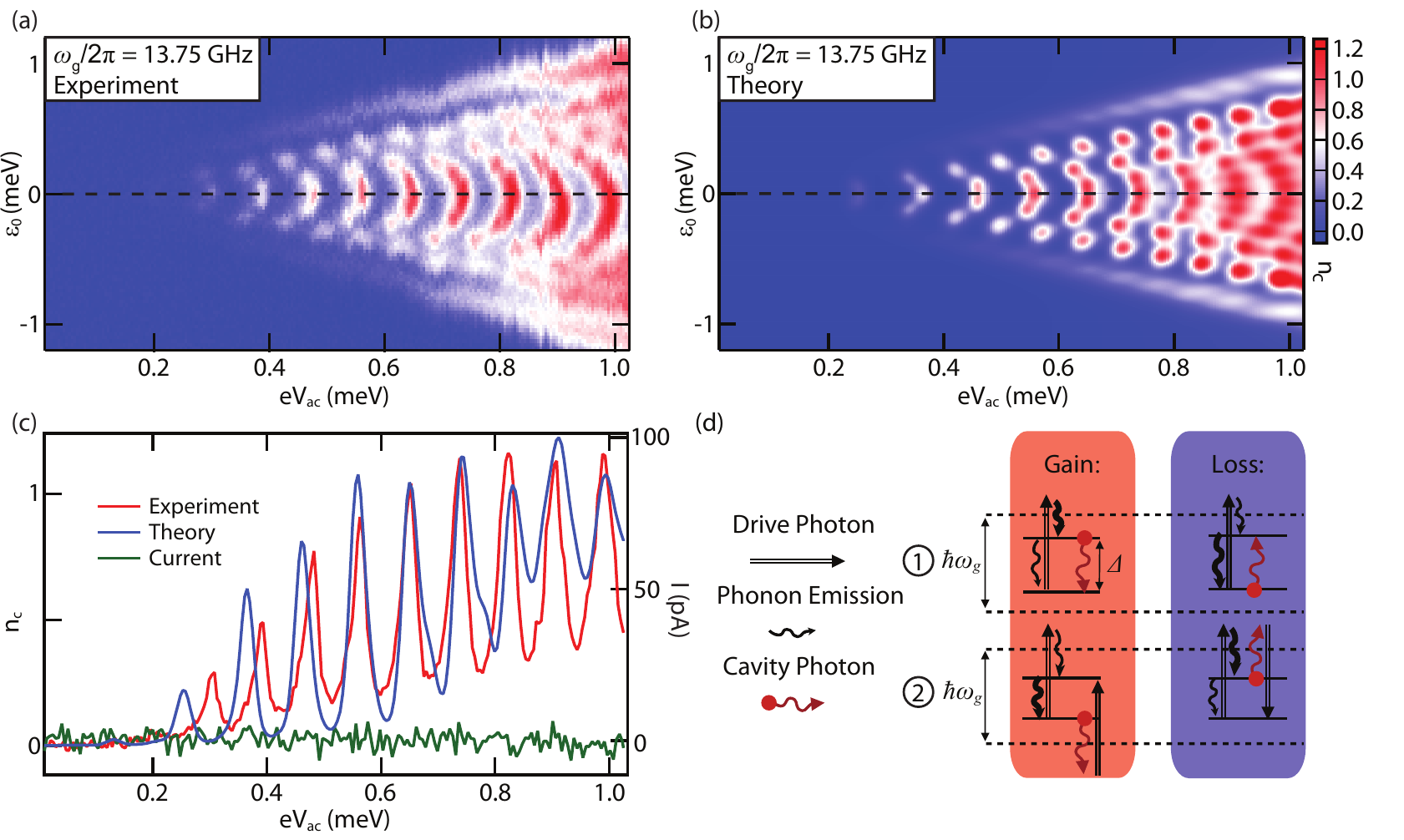}
\caption{(a) Photon number interference. $n_{\rm c}$ measured as a function of $\epsilon_0$ and $e V_{\rm ac}$ with no cavity drive.  (b) Calculated $n_{\rm c}$ as a function of $\epsilon_0$ and $e V_{\rm ac}$, showing good qualitative agreement with the data. (c) Measured and calculated $n_{\rm c}$ as a function of $e V_{\rm ac}$ for $\epsilon_0 = 0$.  Overlaid is the measured current $I$ through the DQD.  Notice that while photons are being emitted, no measurable current is seen. (d) Energy level diagrams for emission and absorption peaks.  Population inversion and emission occur when the quasi-energy splitting $\Delta = \hbar \omega_{\rm c}$ and the DQD can either directly emit a cavity photon [labeled as (1)], or when or $\Delta = \hbar \omega_{\rm g} - \hbar \omega_{\rm c}$ and the DQD can simultaneously absorb a drive photon and emit a cavity photon [labeled as (2)]. The equilibrium population is determined by a competition between driven and undriven phonon relaxation, in the schematic we indicate the dominant phonon relaxation process by bolder arrows.} 
\label{fig:sisyphus:emission}
\end{center}
\vspace{-0.6cm}
\end{figure*}

We now show that the strongly-driven DQD emits light into the cavity mode in the absence of an input field. The intra-cavity photon number $n_{\rm c}$ is plotted as a function of $\epsilon_0$ and $e V_{\rm ac}$ in Fig.\ \ref{fig:sisyphus:emission}(a). These data are obtained by measuring the cavity output power in a 2.5 MHz band around the cavity center frequency (see Appendix \ref{appendix:methods:amp}). The data feature oscillations in $n_{\rm c}$ with a periodicity of approximately 100 $\mu$eV in the drive amplitude $e V_{\rm ac}$.  Oscillations as a function of $\epsilon_0$ are also evident, with up to 7 peaks visible at $e V_{\rm ac} = 0.8$ meV. The locations of the maxima and minima in $n_{\rm c}$ are aligned with the maxima and minima observed in the cavity power gain $G$ [see Fig.\ \ref{fig:sisyphus:transmission}(a)]. We note that Fig.\ \ref{fig:sisyphus:emission}(a) features a slowly rising background, which we attribute to heating due to the strong microwave drive (see Appendix \ref{appendix:theory}).

While the device emits photons in wide range of parameter space, there is no detectable current flow through the device [see Fig.\ \ref{fig:sisyphus:emission}(c)]. The lack of current supports the two-level approximation made in defining the DQD Hamiltonian. It also suggests that the emission can be interpreted as resulting from a single electron being repeatedly forced into the excited state only to decay by producing a cavity photon.  Such processes are often referred to as Sisyphus pumping \cite{grajcar}, as this is analogous to Sisyphus, the character in Greek mythology who eternally pushes a boulder up a hill only for it to roll back down (relax). The experiments presented here are in contrast with previous quantum dot photon emission studies, where a source-drain bias was applied across a DQD to induce an inelastic current flow \cite{YinyuPRL,PhysRevLett.115.046802,PhysRevB.89.165404}.

\section{Charge-cavity coupling in the presence of phonons}

The intracavity photon number $n_{\rm c}$ displays a pattern that is suggestive of Landau-Zener-St\"uckleberg interference \cite{ShevchenkoReview,OliverScience}. However, it is not possible to achieve population inversion in a driven two-level system without including some modification of the dissipation by the drive, as seen in lasing without inversion \cite{AWI} or in Raman phonon emission \cite{inversionReilly}. A qualitative picture of drive assisted thermalization leading to inversion is nevertheless possible.  In Fig.\ \ref{fig:sisyphus:emission}(d) we show energy level diagrams for specific driving conditions that lead to population inversion and large intra-cavity photon numbers.  The inversion, where Floquet quasi-energy eigenstates provide gain for the cavity, arises from a competition between bare phonon relaxation and drive-assisted phonon relaxation, where the system both emits a phonon and absorbs a drive photon. 

To describe this process more quantitatively, we explicitly account for the electron-phonon interaction by adding to Eq.\ (\ref{eqn:hdqd}) the following term:

\begin{equation}
H_{\rm ep} = \sum_{k \nu} \hbar \omega_{k\nu} a_{k\nu}^\dagger a_{k\nu}+\lambda_{k\nu} \sigma_z (a_{k\nu}+a_{k\nu}^\dagger),
\end{equation}
where $k$ refers to the phonon wavevector along the nanowire, $\nu$ indexes the transverse phonon modes, $\omega_{k\nu}$ is the phonon dispersion, $a_{k\nu}~(a_{k\nu}^\dagger)$ is the bosonic annihilation (creation) operator for the phonons, and $\lambda_{k\nu}$ is the electron-phonon coupling constant.   The total Hamiltonian maps to the spin-boson model \cite{Leggett87}, with the caveat that one of the bosonic modes is the cavity field.   In the absence of the drive ($A=0$), the spontaneous phonon relaxation between the two eigenstates of Eq.~(\ref{eqn:hdqd}) is 
\begin{align} \label{eqn:gam}
\gamma(\omega_d) &= 2 \pi  \cos^2\theta  \mathcal{J}(\omega_d),\\
\mathcal{J}(\omega)&=\sum_{k\nu}  |\lambda_{k\nu}|^2 \delta(\omega-\omega_{k\nu}), 
\end{align}
where $\omega_d = \sqrt{\epsilon_0^2+4 t_c^2}$ is the DQD energy splitting, $\theta = \tan^{-1}(2 t_c/\epsilon_0)$ is the mixing angle between the left-right charge states, and $\mathcal{J}(\omega)$ is the phonon spectral density.  
%This approach treats the phonons in a Born-Markov approximation, which is justified when $\gamma \ll \omega_d$.  
The spontaneous emission is accompanied by thermal emission and absorption of the phonons at the  rates $\gamma_{\rm th}(\omega_d)=\gamma(\omega_d) n_p(\omega_d)$, where $n_p(\omega)=(e^{\hbar \omega/k_B T} - 1)^{-1}$ is the Bose-Einstein distribution function at temperature $T$.  In this case, the phonons primarily serve to thermalize the DQD charge states with the lattice.   

In the presence of the drive $(A \ne 0)$ this picture is strongly modified because of the drive-assisted relaxation processes shown in Fig.\ \ref{fig:sisyphus:emission}(d).  In this case, for every positive integer $n$ we need to account for all  phonon emission and absorption processes that  absorb  $n$ photons from the drive.  Each of these processes interact with phonons at a different frequency, which are characterized by the spectral density \cite{gullansSisyphus}
\begin{equation} \label{eqn:Jns}
\begin{split}
\mathcal{J}_{n\pm}(\Delta)& = \sum_{k\nu} |\lambda_{k\nu}|^2 |u_n|^2 \delta(n  \omega \mp \Delta-\omega_{k\nu}) \\
&= {|u_n|^2} \mathcal{J}(n\omega\mp\Delta),
\end{split}
\end{equation} 
where $\pm$ refers to phonon emission processes where the DQD ends in the upper (lower) Floquet state and $u_n$ (found numerically)  accounts for the change of basis to the Floquet eigenstates.  Each of these spectral densities is associated with a spontaneous phonon emission rate $\gamma_{n\pm}(\Delta)= 2\pi \mathcal{J}_{n\pm}(\Delta)$ and thermal emission and absorption rates $\gamma_{n\pm}(\Delta) n_p(n\omega \pm \Delta)$.  Figure \ref{fig:sisyphus:emission}(d) diagrammatically illustrates the two dominant processes at low drive amplitudes: $\gamma_{0-}$ and $\gamma_{1+}$.  The first process $\gamma_{0-}$ corresponds to the bare phonon relaxation at the Floquet quasienergy gap $\Delta$.  The second process $\gamma_{1+}$, on the other hand, describes a process where the DQD absorbs one quanta from the drive and emits a phonon at the energy $\hbar \omega_g - \Delta$, which is greater than zero by the definition of $\Delta$.  We see from Eq.~(6) that these phonon emission rates are sensitive to the frequency dependence of $\mathcal{J}(\omega)$ and the overlap factor $u_n$, the latter of which is a function of the qubit and drive parameters.   This should be contrasted with the standard inclusion of dephasing and decay in the Landau-Zener-St\"uckleberg theory. In this treatment, the decay rates are assumed to be given by their values at $A$  = 0 and never lead to population inversion \cite{ShevchenkoReview}.  Our model, on the other hand, allows for the modification of the spectral density by the drive. As a result, the dominant relaxation process, which determines whether there is population inversion or not, is modulated by the system parameters, leading to the oscillating regions of gain and loss observed in Fig. \ref{fig:sisyphus:transmission}(a).

With this understanding we apply the detailed theory of Ref.~\onlinecite{gullansSisyphus} to make a direct comparison with the experimental data. Quasi-static charge noise and amplitude variations are included in the model by smoothing the calculated cavity response with a Gaussian of width $50\ \mu$eV along the $\epsilon_0$ axis and $10$~$\mu$eV along the $eV_{\rm ac}$ axis. A best fit is obtained with $g_0/2\pi  = 75$ MHz and $t_{\rm c} = 85$~$\mu$eV, both of which are consistent with our measured values \cite{KarlFancy}.  The calculated $n_{\rm c}$ as a function of $\epsilon_0$ and $e V_{\rm ac}$ is shown in Fig.\ \ref{fig:sisyphus:emission}(b) and is in very good agreement with the measured response.  We further compare the theoretical predictions with the experimental results in Fig.\ \ref{fig:sisyphus:emission}(c), where we plot $n_{\rm c}$ as a function of $e V_{\rm ac}$ for $\epsilon_0 = 0$.  Both measured and calculated responses feature strong oscillations with nearly equivalent periodicities.  The discrepancy between the measured and calculated response, especially at high drive amplitudes, may be due to imperfect treatment of the phonon spectrum.

\section{Conclusions and Outlook}

In conclusion, we have demonstrated that a strongly driven semiconductor DQD emits microwave frequency photons, even in the absence of a dc current. Electron-phonon coupling and the periodic detuning drive result in effective population inversion of the Floquet states. The population inversion is responsible for the cavity power gain and photon emission that are observed in our experiments.

Looking beyond the intriguing interference patterns that are observed, the statistics of the light generated by the driven DQD is predicted to be a sensitive function of $\omega_{\rm g}$ and $e V_{\rm ac}$ \cite{gullansSisyphus}.  This means that a single system could span the spectrum of incoherent thermal emission, coherent emision (lasing), and even anti-bunched light by simply changing the drive parameters.  Such tunability is highly desirable and could be used for a detailed study of the transition between classical and non-classical light. 

Our experimental platform may also be useful for spectroscopy. The Floquet energy imbalances that lead to gain and photon emission are very sensitive probes of the spectral properties of the relaxation mechanism of the qubit.  In our case this is the phonon spectral density.  The fine features of the gain plot could therefore be used to further study lifetime limiting mechanisms in InAs and other material systems.  
 
As the technology to scale up solid-state qubit systems continues to develop, it becomes possible to imagine hybrid systems composed of multiple classes of mesoscopic quantum devices on the same chip.  This work illustrates the utility of gate defined quantum dots in such hybrid mesoscopic systems.  Their broad tunability and  unique light-matter interactions, arising from strong electron-phonon coupling, may provide device functionality that is not available in other solid-state qubit systems.

\section{Acknowledgments}

We thank Bert Harrop for help with sample preparation. Work was supported by the Gordon and Betty Moore Foundation's EPiQS Initiative through Grant GBMF4535, with partial support from the National Science Foundation (DMR-1409556 and DMR-1420541). Devices were fabricated in the Princeton University Quantum Device Nanofabrication Laboratory.\\

%%%%%%%%%%%%%%%%%%% Appendix %%%%%%%%%%%%%%%%%%%

\appendix
\section{Experimental Methods}
\label{appendix:methods}

\subsection{Cavity-coupled DQD}
\label{appendix:methods:DQD}

\begin{figure*}
	\begin{center}
		\includegraphics[width=2\columnwidth]{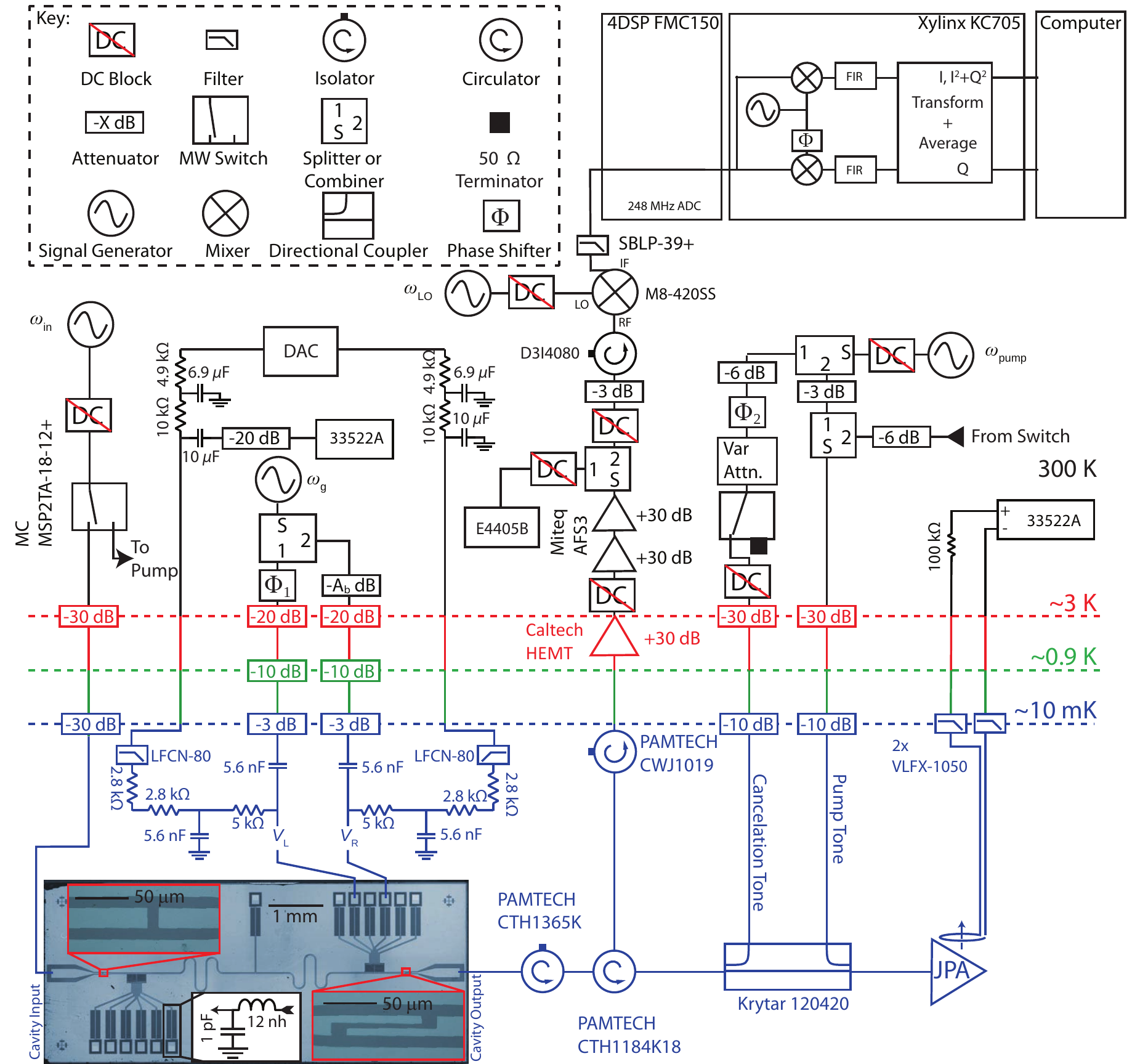} 
		\caption{Complete circuit diagram used to implement the experiment.}
		\label{fig:somsisyphus:setup}
	\end{center}
\end{figure*}

The device consists of a superconducting cavity that contains a single InAs nanowire DQD. An optical micrograph of the sample is shown in the lower left corner of Fig.\ \ref{fig:somsisyphus:setup}. The 50 nm diameter nanowire is placed on top of five predefined Au gate electrodes \cite{FasthInAs,StefanNP,SchroerNano}. The gate electrodes are electrically isolated from the nanowire by a 20 nm layer of silicon nitride $\mathrm{SiN_{x}}$. Source and drain contacts consisting of 20 nm Ti/180 nm Au are defined using electron beam lithography \cite{SulfurEtch}. As in previous experiments, the source contact is directly connected to the center pin of the microwave cavity \cite{KarlFancy}.

The microwave cavity consists of a half-wavelength ($\lambda$/2) superconducting resonator, which is fabricated by selectively etching a 50 nm Nb film that is sputter deposited on an oxidized Si substrate (250 nm) with resistivity $\rho$ $>10$ k$\Omega \cdot$cm. The cavity center frequency is $f_c$ = $\omega_c/2\pi$ = 7.553 GHz and the cavity linewidth is $\kappa/2\pi$ = 1.5 MHz, giving a quality factor $Q \approx 5\,000$.

\label{appendix:methods:amp}

To increase the photon collection efficiency we use asymmetric coupling capacitors with $\kappa_{\rm in}/ 2 \pi = 0.04$ MHz and $\kappa_{\rm out} / 2 \pi  = 0.8$ MHz. Microwave losses were reduced by adding lumped element capacitor-inductor low-pass filters to each gate line. The inductance $L = 12$ nH and capacitance $C = 1$ pF were chosen to give a cutoff frequency around 1.4 GHz, well below $f_c$ = 7.553 GHz. From the measured $Q$   = $\omega_c/(\kappa_{\rm out} + \kappa_{\rm in} + \kappa_{\rm int})$ we estimate the decay rate due to internal losses to be $\kappa_{\rm int}  / 2 \pi \approx 0.6$ MHz, which is an improvement over previous studies \cite{YinyuScience}. 

The charge-cavity coupling strength $g_0$ was increased by reducing the separation between the nanowire source and drain contacts to 390 nm. The smaller lithographic dimensions increased the electric field in the region containing the nanowire DQD.  As a result we achieve $g_0 / 2 \pi \approx 80$ MHz,which is significantly higher than in previous samples ($g_0 / 2 \pi \approx 30$ MHz) \cite{KarlFancy,YinyuScience}.

\subsection{Detuning drive setup}
\label{appendix:methods:detuning}

\begin{figure}
	\begin{center}
		\includegraphics[width=\columnwidth]{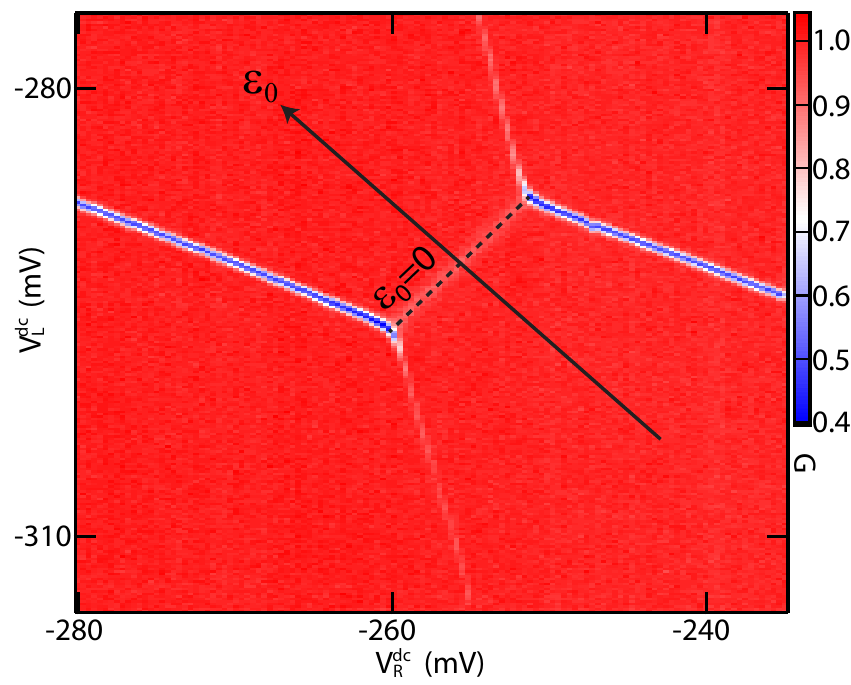}
		\caption{Cavity power gain $G$ plotted as a function of $V_{\rm R}^{\rm dc}$ and $V_{\rm L}^{\rm dc}$ with $e V_{\rm ac}  =0$, showing the charge stability diagram of the DQD. The detuning axis is shown by the solid black arrow. The interdot charge transition, where $\epsilon_0$ = 0, is indicated by the dashed line.} 
		\label{sisyphus:wallwall}
	\end{center}
\end{figure}

The DQD energy level detuning $\epsilon(t)$ is driven by applying time-dependent voltages with frequency $f_g$ = $\omega_{\rm g} / 2 \pi$ to two of the gate electrodes $V_{\rm L}(t)$ and $V_{\rm R}(t)$. The microwave voltages are added to the dc voltages $V_{\rm L}^{\rm dc}$ and $V_{\rm R}^{\rm dc}$, which are used to generate the DQD confinement potential. These dc voltages are generated at room temperature using heavily filtered digital-to-analog converters (DACs).  As result $V_{\rm L}(t)$ and $V_{\rm R}(t)$ become:\begin{equation}
V_{\rm L}(t) = V_{\rm L}^{\rm dc} + V_{\rm L}^{\rm ac}\sin \left( \omega_{\rm g} t \right),
\end{equation}
\begin{equation}
V_{\rm R}(t) = V_{\rm R}^{\rm dc} + V_{\rm R}^{\rm ac} \sin \left( \omega_{\rm g} t + \phi_1  \right).
\end{equation}
Here $V_{\rm L}^{\rm ac}$ and $V_{\rm R}^{\rm ac}$ are the amplitudes of the microwave signals that are applied to the gates. In general these amplitudes are different due to imperfections in the two coaxial lines and the lumped-element $LC$ filters that are located on each gate line. The relative phase $\phi_1$ is due to different coax cable lengths and a variable phase shifter. A fixed attenuator $A_{\rm b}$ = 3 dB is used to balance $V_{\rm R}^{\rm ac}$ and $V_{\rm L}^{\rm ac}$ By adjusting the phase shifter such that $\phi_1 \approx \pi$, we achieve driving that is primarily along the interdot detuning axis shown in Fig.\ \ref{sisyphus:wallwall}.
 
\subsection{Amplification chain and calibration of the intra-cavity photon number}
\label{appendix:methods:amp}

To detect the relatively weak photon emission from the cavity we employ a Josephson parametric amplifier (JPA) \cite{Squeezed1,CavesPRD1982,CavesPRA2012,LehnertJPA,PhysRevApplied.4.014018}.  The cavity output is first passed through an isolator and a circulator before being amplified by the JPA. The signal is further amplified by a high electron mobility transistor (HEMT) amplifier with a 4 Kelvin noise temperature and two Miteq room temperature amplifiers. The JPA is driven by a strong microwave pump tone at frequency $\omega_{\rm pump} / 2\pi = 7.54776$ GHz. The pump tone is added to the cavity output signal using a Krytar directional coupler. Due to imperfect isolation, a small portion of the pump tone may reflect off of the JPA and reach the output port of the cavity. The Krytar directional coupler is also used to couple a cancellation tone to the cavity output port. The cancellation tone originates from the same signal generator, but it is attenuated and phase shifted to result in nearly perfect destructive interference of the reflected pump tone. The cancel tone can be turned off via a room-temperature microwave switch.

After being amplified the signal is either measured using an Agilent E4405B spectrum analyzer or downconverted to 20.48 MHz using a Marki M8-420SS mixer (with a local oscillator $\omega_{\rm LO} / 2 \pi = 7.574295$ GHz). The downconverted signal is digitized by a 4DSP FMC150 fast ADC board and processed using a Xylinx KC705 FPGA board.  Our FPGA program first generates the $I$ ($Q$) quadratures by multiplying the signal with a 20.48 MHz sine (cosine) wave.  The output is then filtered using a 200 element finite impulse response (FIR) filter with a 2.5 MHz bandwidth.  The resulting signal is either kept as $I$ and $Q$ streams and averaged or $I^2 +Q^2$ is calculated and then averaged.  For gain measurements we measure the time averaged quadratures $\{ \overline{I},\overline{Q} \}$.  We can extract the power gain $G = C P_{\rm out}/ P_{\rm in}$ by noting its relationship with the transmitted amplitude $T$:
\begin{equation}
G = C \left| T \right|^2 = C \left| \overline{I} + i \overline{Q} \right|^2.
\end{equation}
As in previous experiments, the normalization constant $C$ is defined such that $G$ = 1 with the device configured in Coulomb blockade \cite{KarlFancy}.

\begin{figure}
	\begin{center}
		\includegraphics[width=\columnwidth]{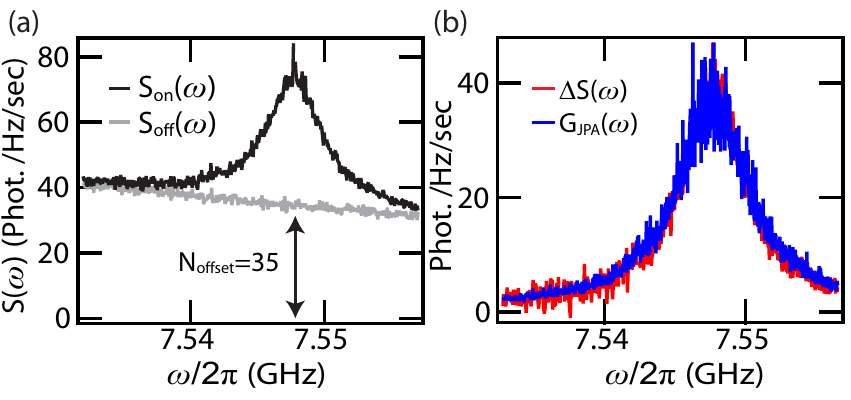}
		\caption{(a) Measured power spectra $S_{\rm on}$ ($S_{\rm off}$) taken with the JPA on (off). The peak in the power spectrum acquired with the JPA turned on corresponds to amplified vacuum noise. The noise offset $N_{\rm offset}$ is referenced to the input of the JPA. (b)  JPA gain $G_{\rm JPA}(\omega)$ overlapped with the difference between the two power spectra $\Delta S(\omega) = S_{\rm on}(\omega) - S_{\rm off}(\omega)$. }
		\label{nofffit}
	\end{center}
\end{figure}

For photon number measurements we measure $\{ \overline{I^2+Q^2}\}$, which is equivalent to measuring the power in the FIR defined band.  We note that this power is proportional to the intracavity photon number $n_{\rm c}$, up to some constant noise offset $N_{\rm offset}$.  $N_{\rm offset}$ is determined by first measuring the power spectral density at the output of the detection chain using an Agilent E4405B spectrum analyzer.  When the parametric amplifier is turned off we measure $S_{\rm off}(\omega)$.  The power that we obtain is proportional to  $N_{\rm offset}$.  Here $N_{\rm offset}$ refers to noise photon flux per unit bandwidth referenced to the input of the JPA.    When the parametric amplifier is turned on, measurements of the power spectral density $S_{\rm on}(\omega)$ exhibit an additional noise contribution with a Lorentzian shape.  We identify the additional noise contribution as amplified vacuum noise.  As a result, when referenced back to the input of the JPA, this contribution will be equivalent to a photon flux per unit bandwidth of $G_{\rm JPA}(\omega) -1$, where $G_{\rm JPA}(\omega)$ is the JPA gain \cite{EichlerPRL2011}.  

We use network analyzer $S$ parameter measurements to independently determine  $G_{\rm JPA}(\omega)$ [see Fig.\ \ref{nofffit}(b)].  We can now scale $\Delta S(\omega) = S_{\rm on} (\omega)- S_{\rm off}(\omega)$ to match $G_{\rm JPA}(\omega) -1$, which fixes the proportionality constant between the absolute measured power and photon flux per unit bandwidth referenced to the input of the JPA. The noise offset in close vicinity to the JPA pump frequency is extracted to be $N_{\rm offset} = 35$ [see Fig.\ \ref{nofffit}(a)] and is attributed to contributions from (i) the HEMT amplifier noise, (ii) losses between the JPA and the HEMT, and (iii) additional noise and losses following the HEMT amplifier.

Knowledge of this noise offset allows us to calibrate the photon flux that arrives at the JPA during cavity emission.  We choose the driving conditions $\epsilon_0 = 0$ and $e V_{\rm ac} = 0.82$ meV (corresponding to a peak in the cavity emission) to generate a steady photon population $n_{\rm c}$ in the cavity.  The total photon flux at the input of the JPA is then $n_{\rm JPA,in} = \eta \kappa_{\rm out} n_{\rm c}$, where $\eta$ = -3.3 dB = 0.46 is the measured transmission coefficient from the cavity to the JPA.  We measure the power spectrum $S_{\rm out}(\omega)$ that results from the cavity emission. Note that for this calibration measurement we turned off the JPA to avoid the influence of non-trivial filtering effects during the analysis.

Based on this measured power spectrum we evaluate the intra-cavity photon number as
\begin{equation}
n_{\rm c} = \frac{1}{\eta} \frac{\kappa}{\kappa_{\rm out}}\int \frac{d\omega}{2 \pi}S_{\rm out}(\omega) =1.1 \pm 0.4.
\end{equation}
Here the uncertainty is due to (i) uncertainty in the value of $\eta$, which is measured at room temperature, (ii) lithographic errors in the cavity coupling capacitor dimensions (which influence $\kappa_{\rm in}$ and $\kappa_{\rm out}$), and (iii) statistical uncertainty in $S_{\rm out}(\omega)$.
We use this knowledge to appropriately scale the photon emission data shown in Fig.\  3(a). 

\section{Gain Video}

\begin{figure}
	\begin{center}
		\includegraphics[width=\columnwidth]{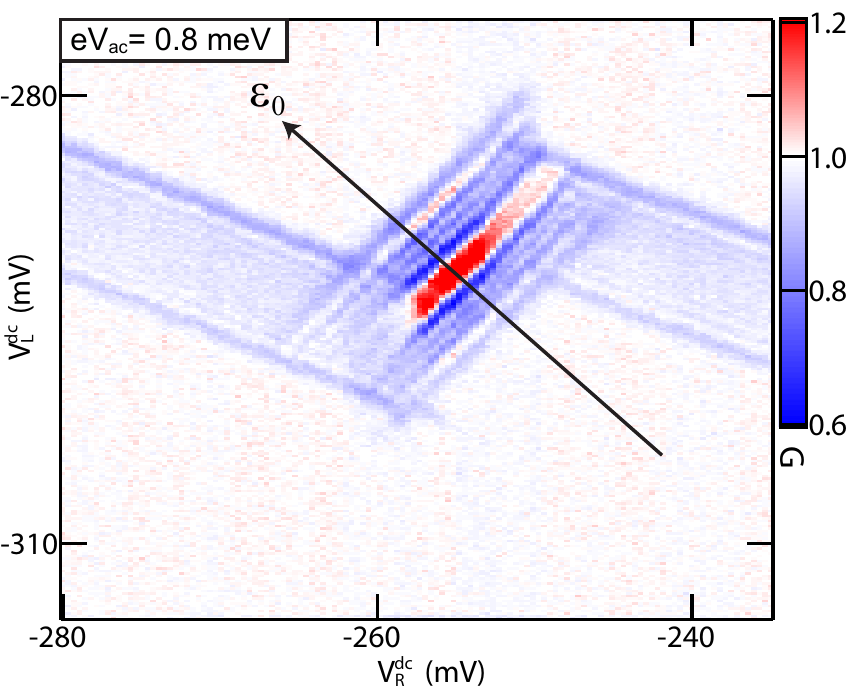}
		\caption{$G$ plotted as a function of $V_{\rm L}^{\rm dc}$ and $V_{\rm R}^{\rm dc}$ for $e V_{\rm ac} = 0.8$ meV.  The plot features strong oscillations parallel to the interdot charge transition line.}
		\label{sisyphus:singleFrame}
	\end{center}
\end{figure}

To further explore the gain medium, we plot $G$ as a function of $V_{\rm L}^{\rm dc}$ and $V_{\rm R}^{\rm dc}$ for $e V_{\rm ac} = 0.8$ meV in Fig.\ \ref{sisyphus:singleFrame}.  The plot features strong oscillations in $G$.  Crucially the gain and loss fringes are parallel to the interdot detuning line.  This signifies that the observed dynamics are independent of the biasing of the DQD energy levels with respect to leads and thus strengthens the interpretation that the observed behavior is a purely two-level system phenomenon.  The fringes weaken near the triple-points of the DQD, which we attribute to photon-assisted-tunneling processes with the leads \cite{RevModPhys.75.1}.  Our detuning axis purposefully avoids these regions. The evolution of the interference pattern is directly visualized in Supplemental Movie I, where we plot $G$ as a function of $V_{\rm L}^{\rm dc}$ and $V_{\rm R}^{\rm dc}$ for increasing $e V_{\rm ac}$.

\section{Theoretical Model}
\label{appendix:theory}

To model the periodically driven DQD we use the Hamiltonian from Eq.~(3)--(4) along with the theoretical methods described in our previous work \cite{gullansSisyphus}.  We move to a Floquet picture where $H(t)$ is mapped to a time-independent Hamiltonian $H_F$ by  
adding the term $\hbar \,\omega \hat{N} =\sum_m \hbar\, \omega\, m \ket{m}\bra{m}$, where $m$ is the Floquet index, and 
converting the functions $e^{-i n \omega t}$  into  operators  which change the Floquet index by $n$ \cite{PhysRevA.7.2203,Grifoni}.  We first find the steady state of the DQD with a finite drive amplitude by neglecting the coupling to the cavity and solving for the dynamics induced solely by phonon relaxation and absorption.  This approach allows us to calculate the steady state population inversion between the two quasi-energy states of the DQD, shown in Fig.\ 3(d), as well as all multi-time DQD correlation functions [e.g., $\langle{\sigma_+(t)\sigma_-(t')}\rangle ]$.
We then use the steady state populations and correlation functions to find the photon correlation function $n_c = \langle{a^\dagger(t) a(t)}\rangle$ \cite{gullansSisyphus}.    All these calculations are done assuming the phonons are at a temperature $T$, which varies linearly with the drive amplitude due to microwave heating of the sample. More precisely $T(eV_\textrm{ac}) = T_0 +  eV_\textrm{ac} \alpha_T$ with $T_0 = 0.02~$K the base temperature of the dilution refrigerator and $\alpha_T = 0.2(1)~$K/meV is chosen to match the oscillation minima in $n_{\rm c}$  [see Fig.\ 3(c)].  

The primary uncertainty in the theoretical model is the precise form for the phonon spectral density $\mathcal{J}(\nu)$.   At these energy scales $J(\nu)$ is dominated by piezoelectric phonons.  The separation between the two dots is $d=100$ nm, the longitudinal confinement is $a=25$ nm for each dot, the phonon speed of sound in the nanowire and substrate is $c_n$ = $4\,000$ m/s and $c_s$ = $11\,000$ m/s, respectively, and the DQD relaxation rate at zero detuning (extracted from current measurements) is $\gamma / 2 \pi=0.5$ GHz \cite{Brandes05,Weber10}.   At large drive amplitudes, virtual processes involving many quanta of the drive are important, which probe $\mathcal{J}(\nu)$ at high frequencies.  As a result, accurate modeling of the data requires a careful treatment of the high frequency behavior of $\mathcal{J}(\nu)$.  We account for this behavior by including, in addition to the coupling to the lowest acoustic branch of the nanowire phonons, coupling of the DQD to acoustic phonons in the substrate \cite{Weber10}.  This treatment gives rise to the functional form for $\mathcal{J}(\nu)$:
\begin{align}
\mathcal{J}(\nu) = &  j_n  \frac{\sin^2 ({ \nu\, d}/{2 c_n}) }{(\nu\,  d/ 2 c_n)} e^{ - \nu a/ c_n} + \nonumber \\
  & j_s \frac{ \nu}{2 c_s/d} [1- \mathrm{sinc}( \nu d/ 2 c_s)] e^{ - \nu a/ c_s},
\end{align}
%\begin{widetext}
%\begin{equation}
%J(\nu) = j_n \frac{\sin^2 ({ \nu\, d}/{2 c_n}) }{(\nu\,  d/ 2 c_n)} e^{ - \nu a/ c_n} + j_s \frac{ \nu}{2 c_s/d} [1- \mathrm{sinc}( \nu d/ 2 c_s)] e^{ - \nu a/ c_s},
%\end{equation}
%\end{widetext}
where the first term arises from coupling to the lowest acoustic phonon branch in the nanowire and the second term accounts for the coupling to acoustic phonons in the substrate (SiN$_x$).  The parameters $j_s$ and $j_n$ are determined by the constraint $\gamma = 2 \pi \mathcal{J}(2 t_c/\hbar) $, with the ratio $r = j_s / j_n $  treated as a free parameter when we fit the data in Fig.~3.  We find the best fit when $r=6$, with consistent results when $r$ is varied up to 50$\%$ of this value.

\bibliographystyle{jason}
\bibliography{GeorgePRX}

\end{document}